# SAFE: Formal Specification and Implementation of a Scalable Analysis Framework for ECMAScript

Hongki Lee
KAIST
petitkan@kaist.ac.kr

Sooncheol Won
KAIST
wonsch@kaist.ac.kr

Joonho Jin
KAIST
myfriend12@kaist.ac.kr

Junhee Cho
KAIST
ssaljalu@kaist.ac.kr

Sukyoung Ryu
KAIST
sryu.cs@kaist.ac.kr

## Abstract

The prevalent uses of JavaScript in web programming have revealed security vulnerability issues of JavaScript applications, which emphasizes the need for JavaScript analyzers to detect such issues. Recently, researchers have proposed several analyzers of JavaScript programs and some web service companies have developed various JavaScript engines. However, unfortunately, most of the tools are not documented well, thus it is very hard to understand and modify them. Or, such tools are often not open to the public.

In this paper, we present formal specification and implementation of SAFE, a scalable analysis framework for ECMAScript, developed for the JavaScript research community. This is the very first attempt to provide both formal specification and its open-source implementation for JavaScript, compared to the existing approaches focused on only one of them. To make it more amenable for other researchers to use our framework, we formally define three kinds of intermediate representations for JavaScript used in the framework, and we provide formal specifications of translations between them. To be adaptable for adventurous future research including modifications in the original JavaScript syntax, we actively use open-source tools to automatically generate parsers and some intermediate representations. To support a variety of program analyses in various compilation phases, we design the framework to be as flexible, scalable, and pluggable as possible. Finally, our framework is publicly available, and some collaborative research using the framework are in progress.

## 1. Introduction

JavaScript is now the language of choice for client-side web programming, which enables dynamic interactions between users and web pages. By embedding JavaScript code that use event handlers such as `onMouseOver` and `onClick`, static HTML web pages become "Dynamic HTML" [12] web pages. JavaScript is originally developed in Netscape, released in the Netscape Navigator 2.0 browser under the name LiveScript in September 1995, and renamed as JavaScript in December 1995. After Microsoft releases its own implementation of the language, JScript, in the Internet Explorer 3.0 browser in 1996, Ecma International develops the standardized version of the language named ECMAScript [8, 9]. JavaScript was first envisioned as a simple scripting language, but with the advent of Dynamic HTML, Web 2.0 [28], and most recently HTML5 [1], JavaScript is now being used on a much larger scale than intended. All the top 100 most popular web sites according to the Alexa list [2] use JavaScript and its use outside web pages is rapidly growing.

```
1  function Wheel4() { this.wheel = 4 }
2  function Car() { this.maxspeed = 200 }
3  Car.prototype = new Wheel4;
4  var modernCar = new Car;
5
6  var beforeModern =
7      modernCar instanceof Car; // true
8
9  function Wheel6() { this.wheel = 6 }
10 Car.prototype = new Wheel6;
11 var afterModern =
12     modernCar instanceof Car; // false
13 var truck = new Car;
14 var aftertruck =
15     truck instanceof Car; // true
```

**Figure 1.** Unintuitive behavior of JavaScript prototypes

As Brendan Eich, the inventor of JavaScript, says [7]:

> "Dynamic languages are popular in large part because programmers can keep types latent in the code, with type checking done *imperfectly* (yet often more quickly and expressively) in the programmers' heads and unit tests, and therefore programmers can do more with less code writing in a dynamic language than they could using a static language."

By sacrificing strong static checking, JavaScript enjoys aggressively dynamic features such as run-time code generation using `eval` and dynamic scoping using `with`. In addition, JavaScript provides quite different semantics from conventional programming languages like C [22] and Java [4]. For example, JavaScript allows programmers to use variables and functions before defining them, and to assign values to new properties of an object even before declaring them in the object. Also, JavaScript allows users to access the global object of a web page via interactions with the DOM (Document Object Model) without requiring any permissions. JavaScript provides "prototype-based" inheritance instead of classes.

Consider the code example in Figure 1. Unlike conventional programming languages, the inheritance hierarchy may be changed after creation of objects. When `modernCar` is constructed at line 4, it is an instance of the `Car` object. However, because the prototype of `Car` is changed from `Wheel4` to `Wheel6` at line 10, `modernCar` is not an instance of `Car` any more at line 12. In JavaScript, when some properties of a constructor change, the objects constructed before and after the change may be considered different instances even though they are constructed from the same constructor.

Due to such quirky semantics, understanding and analyzing JavaScript applications are well known to be difficult, and they are often targets of security attacks [19]. Because of the crude control by the same-origin policy of HTML, once a web page trusts third-party code it permits subsequent contents from the same origin, which often allows malicious scripts to sneak in. Such code injections can easily allow attackers to get high access permissions to secure contents including session cookies and unprotected personal information. This security problem known as XSS (cross-site scripting) shows up often in web pages and web applications. To resolve the problem, web service companies have developed several defense mechanisms such as cookie-based security policies and filtering out string inputs that may contain malicious scripts, but their functionalities are very limited. More robust approach might be using a presumably *safe subset* of JavaScript: Yahoo! ADsafe [6], Facebook FBJS [10], and Google Caja [13]. While they are intended to be safe subsets of JavaScript, none of them has been shown safe. Rather, researchers have reported security vulnerabilities with ADsafe and FBJS [24, 25, 31].

Clearly, better analyses of JavaScript applications for developing more reliable programs become indispensable. As more fundamental solutions to the security vulnerability problems of JavaScript, researchers recently have proposed formal specifications [11, 17, 23], type systems [3, 18, 33], static analyses [16, 20], and combinations of static and dynamic analyses [5, 24] for JavaScript. Web-service companies also have fertilized the JavaScript research community by open sourcing their JavaScript engines such as Rhino [26] and SpiderMonkey [27] from Mozilla, and V8 [14] from Google. While each of them contributes various aspects to solve the problem of JavaScript vulnerability issues, they are yet unsatisfactory in several reasons. First, most of them do not have a well-defined specification or a document to describe them; it is hard for other researchers to understand them and utilize them for their own further research. Secondly, they are not designed and developed for general research but often tightly coupled with their underlying browsers; it is quite challenging to integrate new ideas and new analyzers to existing systems. Thirdly, it is almost impossible to change or extend the existing implementations: most of them do not have any implementations yet, they do not make the implementations available to the public, or the design of the hand-written parsers and Abstract Syntax Tree (AST) nodes are not well-suited to extension and experimentation for researchers since they are with full of undocumented optimizations. Finally, even though the 5th edition of ECMA-262 [9] is released in 2009, most of them deal with the 3rd edition of ECMA-262 [8] released in 1999.

In this paper, we present formal specification and implementation of SAFE (Scalable Analysis Framework for ECMAScript) [30], developed for the JavaScript research community. Based on our own struggles and experiences, the first principles of our framework are formal specification, flexible, scalable, and pluggable framework design, open-source implementation, and aggressive use of various tools for automatic generation. Unlike most of the existing approaches, our framework deals with the 5th edition of ECMA-262 (hereafter called the ECMAScript specification). To help other researchers to understand our framework more easily, we specify every intermediate representation used in the framework formally, and we try to narrow the gaps between the specification and the corresponding implementation. To allow adventurous research ideas to be realized on top of our framework, we use automatically generated parsers and AST nodes from high-level, brief descriptions thanks to various third-party open-source tools. To support a variety of analyses on various compilation phases, we provide three levels of intermediate representations and well-defined translation mechanisms between them. Using SAFE, some collaborative research on JavaScript such as clone detection and code structure analysis are in progress with both academia and industry.

In short, our contributions are as follows:

- SAFE is the very first attempt to support both formal specification and its implementation for JavaScript.
- SAFE is based on the 5th edition of the ECMAScript specification.
- SAFE formally defines every intermediate representation used in the framework and provides formal specifications of the translations between them.
- SAFE describes the formal semantics of its Intermediate Representation (IR) with the descriptions of the corresponding language constructs in the ECMAScript specification.
- SAFE consists of formally defined components that are adequate for pluggable analysis extensions.
- SAFE makes its implementation available to the public for the research community:

  `http://plrg.kaist.ac.kr/research/safe`

## 2. SAFE

Before describing the formal specification and the implementation of SAFE in detail in Sections 3 and 4, we describe the motivation of our work and a big picture of the framework.

### 2.1 Motivation

We encountered several obstacles while using existing tools in our previous research. Recently, we have worked on JavaScript-related topics: 1) adding modules to the existing JavaScript language via desugaring [21] and 2) removing the `with` statement in JavaScript applications [29]. For 1), we designed a module system for JavaScript and devised a desugaring mechanism from JavaScript extended with the module system to a slightly modified $\lambda_{JS}$ [17]. Following the tradition of $\lambda_{JS}$, we extended the implementation of $\lambda_{JS}$ and its desugaring mechanism to handle our module system. We have been very grateful for the authors to open source their implementation but the paper does not describe the desugaring process in detail, the implementation in multiple languages including Haskell and Scheme is not well documented, and the big semantic gap between JavaScript and $\lambda_{JS}$ is not helpful to reason about the original JavaScript applications. For 2), we tried three open-source JavaScript parsers and engines: PluginForJS[1] in C#, Caja[2] in Java, and Rhino in Java. PluginForJS does not cover the entire JavaScript language, Caja supports a dialect of JavaScript, and Rhino uses a hand-written parser with undocumented optimizations and a set of simplified AST nodes. Finally, all of them deal with the 3rd edition instead of the 5th edition of the ECMAScript specification.

Based on our own struggles, we design and develop SAFE, a scalable ECMAScript analysis framework for the JavaScript research community. We present formal specifications of intermediate representations and translations between them for other researchers to understand our framework as easily and quickly as possible. Many parts of the formal specifications of SAFE describe the corresponding sections in the ECMAScript specification to help the readers to consult with the specification. To allow aggressive modifications even on the syntax of JavaScript, we actively use automated generation tools such as Rats! [15] for parsers and ASTGen [32] for intermediate representations. We make our framework open to the public so that any JavaScript research groups can save their work on developing a series of routine compilation phases. At the same time, the framework is modularly designed and developed so that new research ideas can be easily realized and tested by developing a pluggable module on top of our framework.

[1] `https://jslexerparser.codeplex.com`

[2] `http://code.google.com/p/google-caja`

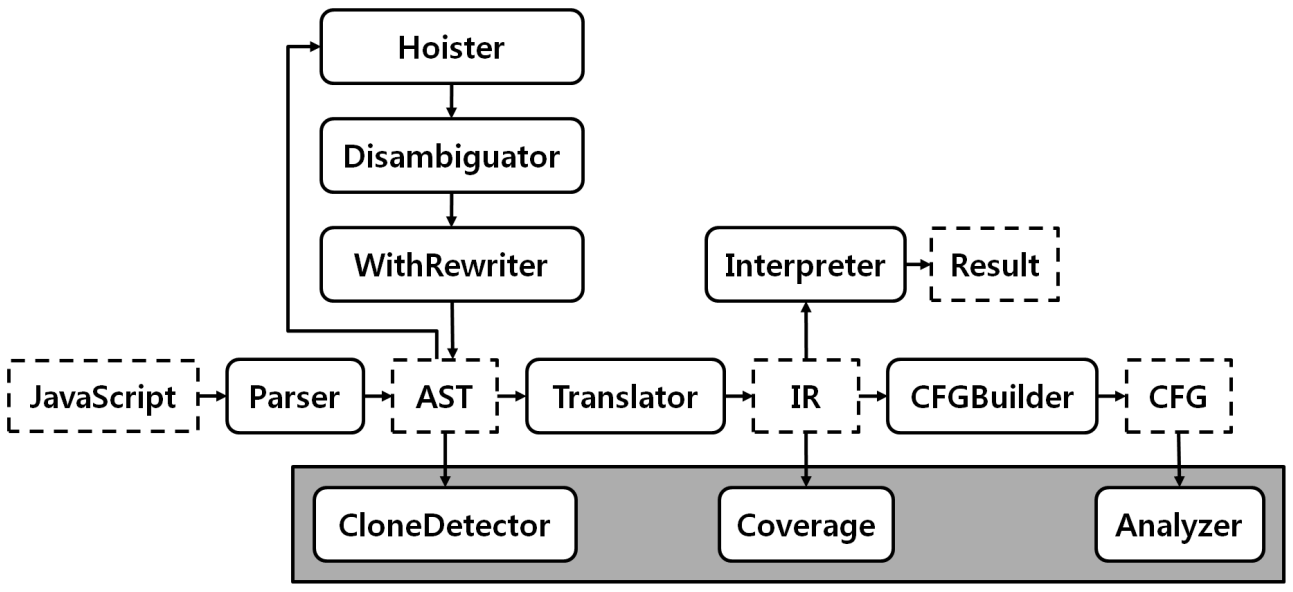

**Figure 2.** SAFE flow graph

### 2.2 Big Picture

Figure 2 describes the overall structure of SAFE. Dashed boxes denote data and solid boxes denote modules that transform data. The framework takes a JavaScript program; `Parser` parses the program and translates it into an AST; a series of compilation phases–`Hoister`, `Disambiguator`, and `WithRewriter`–transforms an AST to a simplified version in AST to make it easier to analyze and evaluate in later phases; `Translator` translates an AST into yet another intermediate representation, Intermediate Representation (IR); finally, `Interpreter` evaluates an IR and produces a result, or `CFGBuilder` constructs a Control Flow Graph (CFG) from an IR to analyze the program. As we describe in later sections, AST, IR, CFG, `Translator`, and `CFGBuilder` are formally specified and their implementations are publicly available.

The shaded box shows additional pluggable components to the framework. Taking advantage of our framework, several collaborative research with academia and industry are in progress: `CloneDetector` detects possible clones among multiple JavaScript applications, `Coverage` calculates the degree to which the JavaScript code has been tested, and `Analyzer` performs a simple type-based analysis of JavaScript programs. Note that each component operates on a different intermediate representation. `CloneDetector` traverses AST nodes, `Coverage` works closely with `Interpreter` on IR, and `Analyzer` scans CFGs for various analyses.

## 3. Formal Specifications

The ECMAScript specification [9] describes the syntax and semantics of JavaScript in prose. The voluminous and informal specification makes it difficult to formally reason about JavaScript applications. While the 258-page specification describes JavaScript in very much detail, it is not rigorous enough: it does not specify every possible case exhaustively, it does not provide a high-level description of various ways to achieve the same behavior, and it includes a plenty of implementation-dependent features. For example, Figure 3 shows the description of the `typeof` operator in the ECMAScript specification, which does not specify the case when evaluating *UnaryExpression* results in an error. Also, JavaScript provides several ways to create function objects, but the specification does not describe them collectively in one place but mentions them sporadically throughout the specification[3]. The underspecified, implementation-dependent, and implementation-defined features result in incompatible JavaScript engines producing different results for the same JavaScript program.

**11.4.3 The `typeof` Operator**

The production *UnaryExpression* : **`typeof`** *UnaryExpression* is evaluated as follows:

1. Let *val* be the result of evaluating *UnaryExpression*.
2. If Type(*val*) is Reference, then
   a. If IsUnresolvableReference(*val*) is **true**, return `"undefined"`.
   b. Let *val* be GetValue(*val*).
3. Return a String determined by Type(*val*) according to Table 20.

**Table 20 — typeof Operator Results**

| *Type of val* | *Result* |
|---|---|
| Undefined | `"undefined"` |
| Null | `"object"` |
| Boolean | `"boolean"` |
| Number | `"number"` |
| String | `"string"` |
| Object (native and does not implement [[Call]]) | `"object"` |
| Object (native or host and does implement [[Call]]) | `"function"` |
| Object (host and does not implement [[Call]]) | Implementation-defined except may not be `"undefined"`, `"boolean"`, `"number"`, or `"string"`. |

**Figure 3.** The `typeof` operator in the ECMAScript specification

In this section, we present the formal specifications of the major components of SAFE.

### 3.1 Intermediate Representations

SAFE provides three levels of intermediate representations: AST, IR, and CFG. The highest level among them is AST, which is very close to the JavaScript concrete syntax; thus, it is the most applicable to source-level analyses such as clone detection. Lower than AST but still higher than machine-level code is IR, which is appropriate for evaluation by an interpreter. IR could be even more compiled down to a lower-level representation for better performance with aggressive code optimizations, and SAFE is open for such a future extension. CFG is the best representation for tracing control flows of a program; most program analyses perform on CFGs. SAFE provides formal specification and implementation of each intermediate representation[4]. Due to space limitations, we describe only IR in this paper and we refer interested readers to the formal specifications of AST and CFG in our open-source repository [30].

Figure 4 presents the syntax of IR. A program $\underline{p}$ in IR is a sequence of IR statements, which consists of function declarations, variable declarations, and IR statements. An IR statement $\underline{s}$ is a simplified version of a corresponding AST statement, and an IR expression $\underline{e}$ is an operator application, a property access, a literal, or an identifier, which does not have any side effects. An IR mem-

[3] The specification describes five ways to create function objects: Section 13.2 describes creating function objects by function declarations and function expressions, Section 15.3.2.1 describes the cases by function constructors as functions and as part of `new` expressions, and Section 15.3.4.5 describes the case by the bind method of function objects.

[4] Formal specifications are available at:
`http://plrg.kaist.ac.kr/redmine/projects/jsf/repository/revisions/master/entry/doc`
and the implementations are available at:
`http://plrg.kaist.ac.kr/redmine/projects/jsf/repository/revisions/master/entry/astgen`
`http://plrg.kaist.ac.kr/redmine/projects/jsf/repository/revisions/master/entry/src/kr/ac/kaist/jsaf/analysis/cfg`

$$
\begin{array}{lcl}
\underline{p} & ::= & \underline{s}^* \\
\underline{s} & ::= & \underline{x} = \underline{e} \\
& | & \underline{x} = \mathsf{delete}\ \underline{x} \\
& | & \underline{x} = \mathsf{delete}\ \underline{x}\texttt{[}\underline{x}\texttt{]} \\
& | & \underline{x} = \texttt{\{}(\underline{m}\texttt{,})^*\texttt{\}} \\
& | & \underline{x} = \texttt{[}(\underline{e}\texttt{,})^*\texttt{]} \\
& | & \underline{x} = \underline{x}\texttt{(}\underline{x}(\texttt{,}\underline{x})^?\texttt{)} \\
& | & \underline{x} = \mathsf{new}\ \underline{x}\texttt{(}(\underline{x}\texttt{,})^*\texttt{)} \\
& | & \underline{x} = \mathsf{function}\ \underline{f}\texttt{(}\underline{x}\texttt{,}\underline{x}\texttt{)}\ \texttt{\{}\underline{s}^*\texttt{\}} \\
& | & \mathsf{function}\ \underline{f}\texttt{(}\underline{x}\texttt{,}\underline{x}\texttt{)}\ \texttt{\{}\underline{s}^*\texttt{\}} \\
& | & \underline{x} = \mathsf{eval}\texttt{(}\underline{e}\texttt{)} \\
& | & \underline{x}\texttt{[}\underline{x}\texttt{]}\ = \underline{e} \\
& | & \mathsf{break}\ \underline{x} \\
& | & \mathsf{return}\ \underline{e}^? \\
& | & \mathsf{with}\ \texttt{(}\underline{x}\texttt{)}\ \underline{s} \\
& | & \underline{x}\ \texttt{:}\ \texttt{\{}\ \underline{s}\ \texttt{\}} \\
& | & \mathsf{var}\ \underline{x} \\
& | & \mathsf{throw}\ \underline{e} \\
& | & \underline{s}^* \\
& | & \mathsf{if}\ \texttt{(}\underline{e}\texttt{)}\ \mathsf{then}\ \underline{s}\ (\mathsf{else}\ \underline{s})^? \\
& | & \mathsf{while}\ \texttt{(}\underline{e}\texttt{)}\ \underline{s} \\
& | & \mathsf{try}\ \texttt{\{}\underline{s}\texttt{\}}\ (\mathsf{catch}\ \texttt{(}\underline{x}\texttt{)}\texttt{\{}\underline{s}\texttt{\}})^?\ (\mathsf{finally}\ \texttt{\{}\underline{s}\texttt{\}})^? \\
& | & \langle \underline{s}^* \rangle \\
\\
\underline{e} & ::= & \underline{e} \otimes \underline{e} \\
& | & \ominus\underline{e} \\
& | & \underline{x}\texttt{[}\underline{e}\texttt{]} \\
& | & \underline{x} \\
& | & \diamond\mathsf{x} \\
& | & \mathsf{this} \\
& | & \underline{\mathit{num}} \\
& | & \underline{\mathit{str}} \\
& | & \mathsf{true} \\
& | & \mathsf{false} \\
& | & \mathsf{undefined} \\
& | & \mathsf{null} \\
\\
\underline{m} & ::= & \underline{x}\ \texttt{:}\ \underline{x} \\
& | & \mathsf{get}\ \underline{f}\texttt{(}\underline{x}\texttt{,}\underline{x}\texttt{)}\ \texttt{\{}\underline{s}^*\texttt{\}} \\
& | & \mathsf{set}\ \underline{f}\texttt{(}\underline{x}\texttt{,}\underline{x}\texttt{)}\ \texttt{\{}\underline{s}^*\texttt{\}} \\
\\
\otimes & ::= & \texttt{|} \mid \texttt{\&} \mid \texttt{\^{}} \mid \texttt{<<} \mid \texttt{>>} \mid \texttt{>>>} \mid \texttt{+} \mid \texttt{-} \mid \texttt{*} \mid \texttt{/} \mid \texttt{\%} \\
& | & \texttt{==} \mid \texttt{!=} \mid \texttt{===} \mid \texttt{!==} \mid \texttt{<} \mid \texttt{>} \mid \texttt{<=} \mid \texttt{>=} \\
& | & \texttt{instanceof} \mid \texttt{in} \\
\\
\ominus & ::= & \texttt{\~{}} \mid \texttt{!} \mid \texttt{+} \mid \texttt{-} \mid \texttt{void} \mid \texttt{typeof}
\end{array}
$$

**Figure 4.** Syntax of the JavaScript IR

ber $\underline{m}$ is either a data property or an accessor property, which is introduced in the 5th edition of the ECMAScript specification. To capture the function call semantics correctly as described in the ECMAScript specification, every function takes exactly two parameters: the first parameter denotes *ThisBinding*, the value associated with the `this` keyword within the function body, and the second parameter denotes an array of the actual arguments.

$$
\begin{array}{rclcl}
(H, A, \mathit{tb}) & \in & \multicolumn{3}{l}{\mathit{Heap} \times \mathit{Env} \times \mathit{ThisBinding}} \\
H & \in & \mathit{Heap} & = & \mathit{Loc} \xrightarrow{\mathit{fin}} \mathit{Object} \\
A & \in & \mathit{Env} & ::= & \#\texttt{Global} \\
& & & | & \mathit{er} :: A \\
\mathit{er} & \in & \mathit{EnvRec} & = & \mathit{DeclEnvRec} \cup \mathit{ObjEnvRec} \\
\sigma & \in & \mathit{DeclEnvRec} & = & \mathit{Var} \xrightarrow{\mathit{fin}} \mathit{StoreValue} \\
l & \in & \mathit{ObjEnvRec} & = & \mathit{Loc} \\
\mathit{tb} & \in & \mathit{ThisBinding} & = & \mathit{Loc}
\end{array}
$$

**Figure 5.** Execution contexts and other domains

$$
\begin{array}{rclcl}
\mathit{ct} & \in & \mathit{Completion} & ::= & \mathit{nc} \\
& & & | & \mathit{ac} \\
\mathit{nc} & \in & \mathit{NormalCompletion} & ::= & \texttt{Normal}(\mathit{vt}) \\
\mathit{ac} & \in & \mathit{AbruptCompletion} & ::= & \texttt{Break}(\mathit{vt}, x) \\
& & & | & \texttt{Return}(v) \\
& & & | & \texttt{Throw}(\mathit{ve}) \\
\mathit{vt} & \in & & & \mathit{Val} \cup \{\texttt{empty}\} \\
\mathit{ve} & \in & \mathit{ValError} & = & \mathit{Val} \cup \mathit{Error}
\end{array}
$$

**Figure 6.** Completion specification type

***Execution Context: Heap, Environment, and ThisBinding*** As the ECMAScript specification describes, when an interpreter evaluates an ECMAScript executable code, it evaluates the code in an *execution context*. We represent an execution context by a triple of a heap, an environment, and a ThisBinding: $(H, A, \mathit{tb})$.

Figure 5 presents a partial set of domain definitions. A heap maps locations to their corresponding objects; an environment is a list of environment records ending with the global object environment record, `#Global`. An environment record is either a declarative environment record or an object environment record: a declarative environment record maps variables to their values, and an object environment record itself is an object. This environment structure is one of the major differences from the 3rd edition of the ECMAScript specification.

***Completion Specification Type*** Under an execution context, evaluating a statement may change the given heap and environment, and it always produces a completion value. As Figure 6 describes, a completion specification type is either a normal completion or an abrupt completion; a normal completion denotes producing a JavaScript value $v$ or nothing (`empty`), and an abrupt completion denotes either diverting the program control via the `break` statement with a value *vt* and a label $x$, returning from a function call with a value $v$, or throwing an exception *ve*. For example, the semantics of the `break` statement is specified as follows:

$$(H, A, \mathit{tb}), \mathsf{break}\ \underline{x} \rightarrow_{\underline{s}} (H, A), \texttt{Break}(\texttt{empty}, \underline{x})$$

Under an executable context $(H, A, \mathit{tb})$, evaluating the `break` statement with a label $\underline{x}$ does not change the heap nor the environment $(H, A)$, and it produces the `Break` completion specification type without any value (`empty`) but with the target label $\underline{x}$.

***Recovering from an Abrupt Completion*** When evaluating a statement results in an abrupt completion, the abrupt completion propagates back to its enclosing statements until a statement recovers the abrupt completion. For example, a `Break` completion with a target label $\underline{x}$ becomes a normal completion when it reaches an enclosing statement labelled with $\underline{x}$:

$$\frac{(H, A, \mathit{tb}), \underline{s} \rightarrow_{\underline{s}} (H', A'), \texttt{Break}(v, \underline{x})}{(H, A, \mathit{tb}), \underline{x}\texttt{:}\ \texttt{\{}\underline{s}\texttt{\}} \rightarrow_{\underline{s}} (H', A'), \texttt{Normal}(v)}$$

When evaluating a statement $\underline{s}$ labelled with a label $\underline{x}$ results in a `Break` completion with a value $v$ and the same label $\underline{x}$, the labelled statement recovers the abrupt completion and produces a normal completion with the value $v$. Similarly, a `Return` completion may become a normal completion by a function call, and a `Throw` completion may become a normal completion by a `try` statement.

***The* `typeof` *Operator*** Now, let us present the operational semantics rules for the `typeof` operator in our IR semantics, which corresponds to the ECMAScript description in Figure 3. Using the following helper function, *TypeTag*, which corresponds to Table 20 in Figure 3:

$$
\mathit{TypeTag}(H, v) \;=\; \begin{cases} \texttt{"undefined"} & \text{if } v \;= \texttt{undefined} \\ \texttt{"object"} & \text{if } v \;= \texttt{null} \\ \texttt{"boolean"} & \text{if } v \in \mathit{Bool} \\ \texttt{"number"} & \text{if } v \in \mathit{Num} \\ \texttt{"string"} & \text{if } v \in \mathit{Str} \\ \texttt{"object"} & \text{if } v \in \mathit{Loc} \land \neg \mathit{IsCallable}(H, v) \\ \texttt{"function"} & \text{if } v \in \mathit{Loc} \land \mathit{IsCallable}(H, v) \end{cases}
$$

we formally specify the operational semantics of the `typeof` operator as follows:

$$
\frac{(H, A, tb), \underline{e} \rightarrow_{\underline{e}} v}{(H, A, tb), \texttt{typeof}\ \underline{e} \rightarrow_{\underline{e}} \mathit{TypeTag}(H, v)}
$$

$$
\frac{(H, A, tb), \underline{e} \rightarrow_{\underline{e}} \mathit{err}}{(H, A, tb), \texttt{typeof}\ \underline{e} \rightarrow_{\underline{e}} \texttt{undefined}}
$$

Unlike the informal description in the ECMAScript specification, our formal specification exhaustively covers all the cases for evaluating the `typeof` operator. The first rule describes that when evaluation of $\underline{e}$ produces a value $v$, evaluation of the `typeof` operator produces a value by using the *TypeTag* helper function. The second rule describes that when evaluation of $\underline{e}$ results in an error, evaluation of the `typeof` operator produces `undefined` as most browsers do.

### 3.2 Translations between Intermediate Representations

In addition to the formal specifications of intermediate representations, SAFE also provides formal specification and implementation of translations between them[5]. Due to space limitations, we describe only several cases of the translation from AST to IR in this paper and we refer interested readers to the formal specification of CFG construction from IR in our open-source repository [30].

Translation from AST to IR consists of translation functions as partially shown in Figure 7. The translation functions maintain an environment $\Sigma$ to handle the names of temporary variables and labels created during translation. Translation of a single AST statement may produce a list of IR statements; we use angle brackets $\langle$ and $\rangle$ to denote a list, and semicolons to denote concatenation of IR statements as a single list. The translation functions use internal names prefixed by $\diamond$: a variable name prefixed by $\diamond$ such as $\underline{\diamond\mathsf{obj}}$ denotes a temporary variable created during translation, and a function name prefixed by $\diamond$ such as $\underline{\diamond\mathsf{getBase}}$ denotes an internal function defined by the IR semantics.

[5] Formal specifications are available at:
`http://plrg.kaist.ac.kr/redmine/projects/jsf/repository/revisions/master/entry/doc`
and the implementations are available at:
`http://plrg.kaist.ac.kr/redmine/projects/jsf/repository/revisions/master/entry/src/kr/ac/kaist/jsaf/compiler`

**12.6 Iteration Statements**

**Syntax**

*IterationStatement* :
    `do` *Statement* `while (` *Expression* `) ;`
    `while (` *Expression* `)` *Statement*
    `for (`*ExpressionNoIn*$_{\mathrm{opt}}$ `;` *Expression*$_{\mathrm{opt}}$ `;` *Expression*$_{\mathrm{opt}}$ `)` *Statement*
    `for ( var` *VariableDeclarationListNoIn* `;` *Expression*$_{\mathrm{opt}}$ `;` *Expression*$_{\mathrm{opt}}$ `)` *Statement*
    `for (` *LeftHandSideExpression* `in` *Expression* `)` *Statement*
    `for ( var` *VariableDeclarationNoIn* `in` *Expression* `)` *Statement*

**Figure 8.** Syntax of iteration statements

The translation function $\mathit{ast2ir}_p[\![p]\!]$ takes a program in AST and produces a program in IR by invoking translation functions on the components of the program: $\mathit{ast2ir}_{fd}[\![fd]\!]$ for function declarations, $\mathit{ast2ir}_{vd}[\![vd]\!]$ for variable declarations, and $\mathit{ast2ir}_s[\![s]\!]$ for statements. As we describe in Section 4.2.1, `Hoister` already reorganized lists of source elements in a program and function bodies in AST so that function declarations and variable declarations appear before statements. Because `Hoister` separates variable declarations from variable initializations, the translation of a variable declaration in AST to IR, $\mathit{ast2ir}_{vd}[\![vd]\!]$, is very simple. The translation of a function declaration is similar to that of a program. The function declaration in IR takes only two parameters; the first parameter denotes the `this` binding for a function call and the second parameter denotes an array of the arguments given at a function call. Accordingly, a function call is translated to take two arguments: the base value of the function reference to denote its `this` binding and an array of the arguments given at the call.

While IR provides a single iteration statement, while, JavaScript supports six statements for iteration as shown in Figure 8: `DoWhile`, `While`, `For`, `ForVar`, `ForIn`, and `ForVarIn`. Among them, `ForVar` and `ForVarIn` are already desugared away by `Hoister` and the others are translated using the IR while statement by `Translator`. The translation of `While` is conventional but that of `ForIn` deserves more attention. Because `ForIn` enumerates the properties of an object and iterates its body until no property remains unvisited, we introduce three internal helper functions: $\diamond$iteratorInit creates an iterator object for a given object, $\diamond$iteratorHasNext checks whether any property remains unvisited, and $\diamond$iteratorNext returns a property name to be visited next. Finally, the translation of a `switch` statement consists of several subsequent translation functions to handle a default clause, if any, and fall through cases.

### 3.3 Example Translation

To illustrate what we have described so far, consider the following JavaScript code:

```
var sum = 0;
for(var i = 1; i <= 10; i++)
  sum+= i;
_<>_print(sum);
```

To show how each part is translated to intermediate representations, we color the corresponding parts in the JavaScript source code and the translated intermediate representations in the same color. Because AST is very similar to the source code, we show only the translated IR and CFG. The code first initializes the variable `sum` to `0` (in orange), and iteratively adds `i` to `sum` (in purple) where `i` is incremented by `1` (in brown) from `1` (in blue) to `10` (in green). To provide a debugging facility for our development, we add a special debugging function $_\diamond_$print. The code ends by printing `sum` (in red).

The following code is a simplified version of the translated IR from the above JavaScript code:

$$\mathit{ast2ir}_p[\![\mathit{fd}^*\ \mathit{vd}^*\ s^*]\!] = \langle(\mathit{ast2ir}_{\mathit{fd}}[\![\mathit{fd}]\!](\langle\rangle))^*\ (\mathit{ast2ir}_{\mathit{vd}}[\![\mathit{vd}]\!]\langle\rangle)^*\ (\mathit{ast2ir}_s[\![s]\!](\langle\rangle))^*\rangle$$

$$\begin{aligned}\mathit{ast2ir}_{\mathit{fd}}[\![\texttt{function}\ f\,\texttt{(}(x\texttt{,})^*\texttt{)}\ \texttt{\{}\mathit{fd}^*\mathit{vd}^*s^*\texttt{\}}]\!](\Sigma) = {} & \mathsf{function}\ \underline{f}\,(\underline{\diamond\mathsf{this}},\ \underline{\diamond\mathsf{arguments}})\{\\ & \quad(\mathit{ast2ir}_{\mathit{fd}}[\![\mathit{fd}]\!](\Sigma))^*\\ & \quad(\mathsf{var}\ \underline{x_i})^*\\ & \quad(\mathit{ast2ir}_{\mathit{vd}}[\![\mathit{vd}]\!]\Sigma)^*\\ & \quad(\underline{x_i} = \diamond\mathsf{arguments}[\texttt{"i"}])^*\\ & \quad(\mathit{ast2ir}_s[\![s]\!](\Sigma;\underline{\diamond\mathsf{this}};\underline{\diamond\mathsf{arguments}}))^*\}\end{aligned}$$

$$\mathit{ast2ir}_{\mathit{vd}}[\![\texttt{var}\ x]\!](\Sigma) = \mathsf{var}\ \underline{x}$$

$$\begin{aligned}\mathit{ast2ir}_{\mathit{lhs}}[\![f\,\texttt{(}(e,)^*\texttt{)}]\!](\Sigma)(\underline{x}) = {} & \text{LET } ((\underline{s}^*,\underline{e}) = \mathit{ast2ir}_e[\![e]\!](\Sigma)(\underline{\diamond\mathsf{y}}))^*\\ & \text{IN}\quad(\underline{\diamond\mathsf{obj}} = \diamond\mathsf{toObject}(\underline{f})\,;\ (\underline{s}^*;\ \underline{\diamond\mathsf{y}} = \underline{e})^*;\\ & \qquad\underline{\diamond\mathsf{arguments}} = [(\diamond\mathsf{y_i},)^*];\\ & \qquad\underline{\diamond\mathsf{fun}} = \diamond\mathsf{getBase}(\underline{f})\,;\\ & \qquad\underline{x} = \underline{\diamond\mathsf{obj}}(\underline{\diamond\mathsf{fun}},\underline{\diamond\mathsf{arguments}}),\underline{x})\end{aligned}$$

$$\begin{aligned}\mathit{ast2ir}_s[\![\texttt{while (}e\texttt{)}\ s]\!](\Sigma) = {} & \text{LET } (\underline{s}^*,\underline{e}) = \mathit{ast2ir}_e[\![e]\!](\Sigma)(\underline{\diamond\mathsf{new}_1})\\ & \text{IN}\quad\langle\underline{\diamond\mathsf{break}} : \{\\ & \qquad\quad\underline{s}^*;\\ & \qquad\quad\mathsf{while}\ (\underline{e})\ \{\\ & \qquad\qquad\underline{\diamond\mathsf{continue}} : \{\mathit{ast2ir}_s[\![s]\!](\Sigma;\underline{\diamond\mathsf{break}};\underline{\diamond\mathsf{continue}})\};\\ & \qquad\qquad\underline{s}^*;\\ & \qquad\quad\}\\ & \qquad\}\rangle\end{aligned}$$

$$\begin{aligned}\mathit{ast2ir}_s[\![\texttt{for (}\mathit{lhs}\ \texttt{in}\ e\texttt{)}\ s]\!](\Sigma) = {} & \text{LET } (\underline{s}^*,\underline{e}) = \mathit{ast2ir}_e[\![e]\!](\Sigma)(\underline{\diamond\mathsf{new}_1})\\ & \text{IN}\quad\langle\underline{\diamond\mathsf{break}} : \{\\ & \qquad\quad\underline{s}^*;\\ & \qquad\quad\underline{\diamond\mathsf{obj}} = \diamond\mathsf{toObject}(\underline{e})\,;\\ & \qquad\quad\underline{\diamond\mathsf{iterator}} = \diamond\mathsf{iteratorInit}(\underline{\diamond\mathsf{obj}});\\ & \qquad\quad\underline{\diamond\mathsf{cond}_1} = \diamond\mathsf{iteratorHasNext}(\underline{\diamond\mathsf{obj}},\underline{\diamond\mathsf{iterator}});\\ & \qquad\quad\mathsf{while}\ (\underline{\diamond\mathsf{cond}_1})\ \{\\ & \qquad\qquad\underline{\diamond\mathsf{key}} = \diamond\mathsf{iteratorNext}(\underline{\diamond\mathsf{obj}},\underline{\diamond\mathsf{iterator}});\\ & \qquad\qquad\mathit{ast2ir}_{\mathit{lval}}[\![\mathit{lhs}]\!](\Sigma)(;\ \underline{\diamond\mathsf{key}})(\mathrm{false})._1;\\ & \qquad\qquad\underline{\diamond\mathsf{continue}}:\{\mathit{ast2ir}_s[\![s]\!](\Sigma;\underline{\diamond\mathsf{break}};\underline{\diamond\mathsf{continue}})\};\\ & \qquad\qquad\underline{\diamond\mathsf{cond}_1} = \diamond\mathsf{iteratorHasNext}(\underline{\diamond\mathsf{obj}},\underline{\diamond\mathsf{iterator}});\\ & \qquad\quad\}\\ & \qquad\}\rangle\end{aligned}$$

$$\begin{aligned}\mathit{ast2ir}_s[\![\texttt{switch (}e\texttt{) \{}cc_1^*\ (\texttt{default:}s^*)^?\ cc_2^*\texttt{\}}]\!](\Sigma) = {} & \text{LET } (\underline{s}^*,\underline{e}) = \mathit{ast2ir}_e[\![e]\!](\Sigma)(\underline{\diamond\mathsf{val}})\\ & \text{IN}\quad\langle\underline{\diamond\mathsf{break}} : \{\\ & \qquad\quad\underline{s}^*;\ \boxed{\underline{\diamond\mathsf{val}} = \underline{e};}\\ & \qquad\quad\mathit{ast2ir}_{\mathit{case}}[\![(\mathsf{rev}\ cc_2^*)(s^*)^?(\mathsf{rev}\ cc_1^*)]\!](\Sigma;\underline{\diamond\mathsf{break}};\underline{\diamond\mathsf{val}})\}\rangle\end{aligned}$$

$$\begin{aligned}\mathit{ast2ir}_{\mathit{case}}[\![(\texttt{case}\ e\ \texttt{:}\ s_1^*) :: cc_2^*\ (s_2^*)^?\ cc_1^*]\!](\Sigma)(c^*) = {} & \langle\underline{\diamond\mathsf{label}} : \{\mathit{ast2ir}_{\mathit{case}}[\![cc_2^*\ (s_2^*)^?\ cc_1^*]\!](\Sigma)((e,\underline{\diamond\mathsf{label}}) :: c^*)\};\\ & \ (\mathit{ast2ir}_s[\![s_1]\!](\Sigma))^*\rangle\end{aligned}$$

$$\begin{aligned}\mathit{ast2ir}_{\mathit{case}}[\![()\ (s^*)^?\ cc_1^*]\!](\Sigma)(c^*) = {} & \langle\underline{\diamond\mathsf{label}} : \{\mathit{ast2ir}_{\mathit{case}}[\![()\ ()\ cc_1^*]\!](\Sigma)(c^*@[((),\underline{\diamond\mathsf{label}})])\};\\ & \ ((\mathit{ast2ir}_s[\![s]\!](\Sigma))^*)^?\rangle\end{aligned}$$

$$\begin{aligned}\mathit{ast2ir}_{\mathit{case}}[\![()\ ()\ (\texttt{case}\ e\ \texttt{:}\ s^*) :: cc_1^*]\!](\Sigma)(c^*) = {} & \langle\underline{\diamond\mathsf{label}} : \{\mathit{ast2ir}_{\mathit{case}}[\![()\ ()\ cc_1^*]\!](\Sigma)((e,\underline{\diamond\mathsf{label}}) :: c^*)\};\\ & \ (\mathit{ast2ir}_s[\![s]\!](\Sigma))^*\rangle\end{aligned}$$

$$\begin{aligned}\mathit{ast2ir}_{\mathit{case}}[\![()\ ()\ ()]\!](\Sigma)((e,\underline{l})^*) = {} & \langle\mathit{ast2ir}_{\mathit{scond}}[\![(e,\underline{l})^*]\!](\Sigma);\\ & \ \mathsf{break}\ \Sigma(\underline{\diamond\mathsf{break}})\rangle\end{aligned}$$

$$\begin{aligned}\mathit{ast2ir}_{\mathit{scond}}[\![(e,\underline{l}) :: (c^*)]\!](\Sigma) = {} & \text{LET } (\underline{s}^*,\underline{e}) = \mathit{ast2ir}_e[\![e]\!](\Sigma)(\underline{\diamond\mathsf{cond}})\\ & \text{IN}\quad\langle\underline{s}^*;\\ & \qquad\mathsf{if}\ (\Sigma(\underline{\diamond\mathsf{val}}) === \underline{e})\ \mathsf{then\ break}\ \underline{l}\ \mathsf{else}\ \mathit{ast2ir}_{\mathit{scond}}[\![c^*]\!](\Sigma)\rangle\end{aligned}$$

$$\mathit{ast2ir}_{\mathit{scond}}[\![[((),\underline{l})]]\!](\Sigma) = \langle\mathsf{break}\ \underline{l}\rangle$$

$$\mathit{ast2ir}_{\mathit{scond}}[\![()]\!](\Sigma) = \langle\rangle$$

*Where* $c$ *is either* $(e,\underline{l})$ *or* $((),\underline{l})$.

**Figure 7.** An excerpt of the translation rules from AST to IR

```
var sum
var i

<>_one_1 = 1
{
  <>_old_2 = sum
  <>_temp_3 = 0
  sum = <>_temp_3
  <>_ignore_4 = sum
}
{
  {
    <>_old_5 = i
    <>_temp_6 = 1
    i = <>_temp_6
    <>_ignore_4 = i
  }
  <>_break_7 : {
    {
      <>_temp_8 = i
      <>_temp_9 = 10
      <>_temp_10 = <>_temp_8 <= <>_temp_9
    }
    <>_temp_11 = <>toBoolean(<>_temp_10)
    while(<>_temp_11)
      {
        <>_temp_12 : {
          <>_old_13 = sum
          <>_temp_14 = i
          <>_new_15 = <>_old_13 + <>_temp_14
          sum = <>_new_15
          <>_ignore_4 = sum
        }
        {
          {
            <>_old_16 = i
            <>_new_17 = <>toNumber(<>_old_16)
            <>_new2_18 = <>_new_17 + <>_one_1
            i = <>_new2_18
            <>_temp_19 = i
          }
          <>_temp_19 = <>_new_17
        }
        {
          <>_temp_8 = i
          <>_temp_9 = 10
          <>_temp_10 = <>_temp_8 <= <>_temp_9
        }
        <>_temp_11 = <>toBoolean(<>_temp_10)
      }
  }
}
{
  <>_temp_20 = sum
  <>_ignore_4 = <>print(<>_temp_20)
}
```

Because the translated IR includes verbose information such as source location, we cleaned up such information to clearly show the correspondence between the original JavaScript source code and the generated IR code. Note that the conditional expression to check whether `i` is less than equal to `10` (in green) shows up twice, before the while statement and inside the while statement.

Now, the following graph presents the generated CFG from the above IR code:

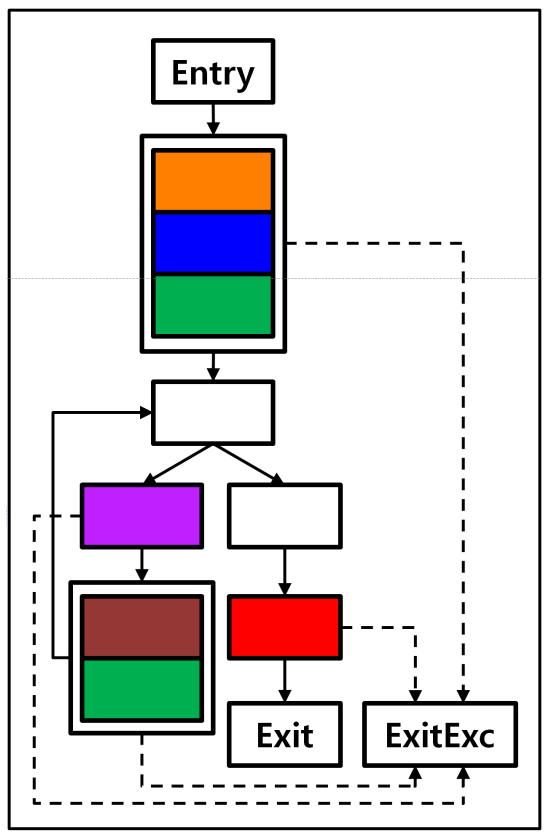

Each colored box denotes a sequence of instructions corresponding to the IR code segment with the same color; several colored boxes constitute a basic block. The `Entry` node denotes the beginning of the program, three consecutive nodes next to `Entry` denote the initialization part of the loop, the left branch with three nodes (of the colors purple, brown, and green) denotes the loop body, and the right branch ending with the `Exit` node denotes the control flow after the loop. Note that the dashed lines to the `ExitExc` node from various basic blocks denote possible exception flows.

## 4. Implementation

In this section, we describe how we realized the formal specifications of SAFE described in the previous section.

### 4.1 Why Yet Another Parser for JavaScript?

As we briefly mentioned in Section 2.1, our previous research on a JavaScript module system required modifications of a JavaScript parser and its AST nodes to extend the syntax to support modules.

```
/**
 * SourceElement ::= Stmt
 */
abstract Stmt();
    /**
     * Stmt ::= do Stmt while ( Expr ) ;
     */
    DoWhile(Stmt body, Expr cond);
    /**
     * Stmt ::= while ( Expr ) Stmt
     */
    While(Expr cond, Stmt body);
    /**
     * Stmt ::= for ( Expr? ; Expr? ; Expr? ) Stmt
     */
    For(Option<Expr> init, Option<Expr> cond,
        Option<Expr> action, Stmt body);
    /**
     * Stmt ::= for ( lhs in Expr ) Stmt
     */
    ForIn(LHS lhs, Expr expr, Stmt body);
    /**
     * Stmt ::= for ( var VarDecl(, VarDecl)* ;
     *                Expr? ; Expr? ) Stmt
     */
    ForVar(List<VarDecl> vars, Option<Expr> cond,
           Option<Expr> action, Stmt body);
    /**
     * Stmt ::= for ( var VarDecl in Expr ) Stmt
     */
    ForVarIn(VarDecl var, Expr expr, Stmt body);
```

**Figure 9.** An excerpt of the high-level AST specification

We considered various parsers and AST structures from academia and industry including ANTLR[6], Scala's parser combinators[7], Rhino, SpiderMonkey, Closure Tools[8], JSConTest[9], and JSure[10], but they were not satisfactory to us. Most of them do not cover the entire JavaScript language and their AST structures do not reflect the JavaScript syntax well. Even though the Rhino parser written in Java is very powerful, because it is a ported version of the hand-written SpiderMonkey parser in C++, we excluded it from consideration for productivity reason.

We actively use open-source tools to automatically generate parsers and intermediate representations. We provide a high-level description of the AST node hierarchy as partially shown in Figure 9 where the indentation denotes a subclassing relationship. Then, ASTGen [32] reads the description and generates Java classes for the AST nodes with some utility methods such as getters, setters, `equals`, and `hashCode`. Similarly, we provide a BNF-style grammar and the corresponding action code for the grammar productions, then Rats! [15] generates a JavaScript parser in Java.

The implementation languages of our framework are Scala and Java, where most of the hand-written code are written in Scala. We use both languages to take advantage of the abundant libraries and tools in Java such as ASTGen and Rats!, and to get benefits from pattern matching and higher-order functions in Scala. Also, because both languages are compiled into Java bytecode, we enjoy the seamless interoperability between them.

---

[6] `http://www.antlr.org`

[7] `http://www.scala-lang.org/api/current/scala/util/parsing/combinator/Parsers.html`

[8] `https://developers.google.com/closure`

[9] `https://github.com/heidegger/JSConTest`

[10] `https://github.com/berke/jsure`

### 4.2 AST Transformations due to the Quirky Semantics

Because of the quirky semantics of JavaScript, SAFE transforms an AST to a simplified version to make it easier to analyze and evaluate in later phases via the following phases:

- `Hoister` lifts the declarations of functions and variables inside programs and functions up to the beginning of them;
- `Disambiguator` checks some static restrictions and renames identifiers to unique names; and
- `WithRewriter` rewrites the `with` statements that do not include any dynamic code generation such as `eval` into other statements without using the `with` statement but with the same semantics.

Because all the intermediate representations preserve backward-mapping information such as source locations, any errors detected on intermediate representations are reported back to the users in terms of JavaScript source locations.

#### 4.2.1 Hoister

When JavaScript code is evaluated, functions and variables declared in the code are first bound in the environment of the running execution context before evaluating the code. In effect, the functions and variables can be used before their textual declarations. For example, the following code:

```
1   x;
2   var x = f();
3   function f() { return 42; }
```

shows that `x` on line 1 is used before its declaration on line 2. Because every compilation phase or an analysis of JavaScript programs should take this feature into account, SAFE *hoists* the declarations of functions and variables to the beginning of the enclosing programs and functions, which preserves the original semantics. By taking care of this feature once at an early phase of compilation, the tasks of the later phases become less burdensome.

After hoisting, the above code is transformed as follows:

```
1   function f() { return 42; }
2   var x;
3   x;
4   x = f();
```

The function declaration is hoisted as it is, but the variable declaration of `x` is split into a declaration without initialization on line 2 and an assignment on line 4.

#### 4.2.2 Disambiguator

Like compilers for conventional programming languages, SAFE renames identifiers in JavaScript code to unique names via the `Disambiguator` phase. By assigning unique names, identifiers with the same name but in different scopes become explicitly distinct, which makes the tasks of the later phases easier. For example, the following code:

```
function f() {
  var x;
  function() { var x; };
}
function h() { var x; }
var x;
```

is transformed as follows:

```
function f() {
  var <>_x_1;
  function() { var <>_x_2; };
}
function h() { var <>_x_3; }
var x;
```

```
abstract class Completion()
case class Normal(v:Option[Val]) extends Completion
abstract class Abrupt() extends Completion
case class Break(v:Option[Val],
                 l:IRId) extends Abrupt
case class Return(v: Val) extends Abrupt
case class Throw(e: ValError) extends Abrupt
```

**Figure 10.** Completion specification type in the interpreter

Also, `Disambiguator` checks some static restrictions such as the following:

- A `return` statement should be within a function body.
- A `continue` statement should occur inside an iterator statement.
- A labelled statement should not be enclosed by another labelled statement with the same label.

#### 4.2.3 `with` Rewriter

The `with` statement in JavaScript makes static analysis of JavaScript applications difficult by introducing a new scope at run time and thus invalidating lexical scoping. Thus, many static approaches to JavaScript program analysis simply disallow the `with` statement. Instead of simply avoiding the uses of the `with` statement, our previous research [29] investigates the usage patterns of the `with` statement in real-world JavaScript applications currently used in the 944 most popular web sites, and it shows that we can rewrite all the static occurrences of the `with` statement that does not have any dynamic code generating functions.

Using the previous research, SAFE provides the `WithRewriter` phase to statically eliminate the `with` statement in AST as long as it is possible. Because evaluating the `with` statement creates an object environment record, the properties of the given `with` object are added to the current lexical environment for the body. Therefore, if the `with` object has a property with the same name as an identifier in the body, the identifier becomes referring to the property; otherwise, it refers to a name bound in its enclosing environment record. The strategy of `with` rewriting is to replace each identifier occurrence in the `with` body with a ternary expression to access the correct entity depending on the existence of a property with the same name in the `with` object. For example, the following `with` statement:

```
with (expr) { a + b; }
```

is rewritten as follows:

```
var $f = toObject(expr);
("a" in $f ? $f.a : a) + ("b" in $f ? $f.b : b);
```

More detailed description of the `with` rewriting is available in our previous work [29].

### 4.3 AST Translations and IR Interpretation

As we discussed in Section 3.2, `Translator` translates an AST into an IR, `CFGBuilder` constructs a CFG from the IR, and `Interpreter` evaluates the IR. Our implementation follows the corresponding formal specifications as closely as possible so that they can be mutually beneficial. For example, Figure 10 presents the corresponding implementation of the completion specification type in Figure 6, which is very close to each other.

**12.6.2 The `while` Statement**

The production *IterationStatement* : **while (** *Expression* **)** *Statement* is evaluated as follows:

1. Let *V* = empty.
2. Repeat
   a. Let *exprRef* be the result of evaluating *Expression*.
   b. If ToBoolean(GetValue(*exprRef*)) is **false**, return (normal, *V*, empty).
   c. Let *stmt* be the result of evaluating *Statement*.
   d. If *stmt*.value is not empty, let *V* = *stmt*.value.
   e. If *stmt*.type is not continue || *stmt*.target is not in the current label set, then
      i. If *stmt*.type is break and *stmt*.target is in the current label set, then
         1. Return (normal, *V*, empty).
      ii. If *stmt* is an abrupt completion, return *stmt*.

**Figure 11.** The `while` statement in the specification

12.6.2 The `while` Statement

$$\frac{(H, A, tb), \underline{e} \rightarrow_{\underline{e}} err}{(H, A, tb), \text{while } (\underline{e})\ \underline{s} \rightarrow_{\underline{s}} (H, A), \texttt{Throw}(err)}$$

$$\frac{(H, A, tb), \underline{e} \rightarrow_{\underline{e}} v \qquad \mathit{ToBoolean}(v) = \texttt{false}}{(H, A, tb), \text{while } (\underline{e})\ \underline{s} \rightarrow_{\underline{s}} (H, A), \texttt{Normal(empty)}}$$

$$\frac{\begin{array}{c}(H, A, tb), \underline{e} \rightarrow_{\underline{e}} v \qquad \mathit{ToBoolean}(v) = \texttt{true} \\ (H, A, tb), \underline{s} \rightarrow_{\underline{s}} (H', A'), ac\end{array}}{(H, A, tb), \text{while } (\underline{e})\ \underline{s} \rightarrow_{\underline{s}} (H', A'), ac}$$

$$\frac{\begin{array}{c}(H, A, tb), \underline{e} \rightarrow_{\underline{e}} v \qquad \mathit{ToBoolean}(v) = \texttt{true} \\ (H, A, tb), \underline{s} \rightarrow_{\underline{s}} (H', A'), nc \\ (H', A', tb), \text{while } (\underline{e})\ \underline{s} \rightarrow_{\underline{s}} (H'', A''), ac\end{array}}{(H, A, tb), \text{while } (\underline{e})\ \underline{s} \rightarrow_{\underline{s}} (H'', A''), ac}$$

$$\frac{\begin{array}{c}(H, A, tb), \underline{e} \rightarrow_{\underline{e}} v \qquad \mathit{ToBoolean}(v) = \texttt{true} \\ (H, A, tb), \underline{s} \rightarrow_{\underline{s}} (H', A'), nc \\ (H', A', tb), \text{while } (\underline{e})\ \underline{s} \rightarrow_{\underline{s}} (H'', A''), \texttt{Normal(empty)}\end{array}}{(H, A, tb), \text{while } (\underline{e})\ \underline{s} \rightarrow_{\underline{s}} (H'', A''), nc}$$

$$\frac{\begin{array}{c}(H, A, tb), \underline{e} \rightarrow_{\underline{e}} v \qquad \mathit{ToBoolean}(v) = \texttt{true} \\ (H, A, tb), \underline{s} \rightarrow_{\underline{s}} (H', A'), nc \\ (H', A', tb), \text{while } (\underline{e})\ \underline{s} \rightarrow_{\underline{s}} (H'', A''), \texttt{Normal}(v)\end{array}}{(H, A, tb), \text{while } (\underline{e})\ \underline{s} \rightarrow_{\underline{s}} (H'', A''), \texttt{Normal}(v)}$$

**Figure 12.** The `while` statement in the IR semantics

By developing both the formal specification and the implementation in parallel, we often found that one form reveals missing cases in the other form, and vice versa. For example, consider the description of the `while` statement in the ECMAScript specification presented in Figure 11. For each construct in the ECMAScript specification, we proceed to describe its operational semantics formally as in Figure 12, and we simultaneously implement its semantics as in Figure 13. The implementation in Scala often matches the specification closely, or the initial implementation looks like a direct translation of the formal specification and we optimize them for performance improvement. In such cases, we keep the original implementation in comment for documentation purpose. While it is often unclear whether the informal prose in the ECMAScript specification covers every possible case, the operational semantics rules are well equipped to cover the cases exhaustively. At the same time, while the operational semantics rules have many duplicated antecedents which make the rules bulky, the concise implementation of the interpreter complements the rules and the pattern matching mechanism of Scala checks the exhaustiveness and redundancy of the cases.

```
/*
 * 12.6.2 The while Statement
 */
case SIRWhile(info, cond, body) =>
  var V: Option[Val] = None
  var AC: Option[Abrupt] = None
  var cont: Boolean = true
  while (cont) {
    walkExpr(cond) match {
      case v:Val => cont = IH.toBoolean(v)
      case err:JSError =>
        AC = Some(throwErr(err, info))
        cont = false
    }
    if (cont) {
      walk(body) match {
        case Normal(v) => if (v.isDefined) V = v
        case ac:Abrupt =>
          AC = Some(ac)
          cont = false
      }
    }
  }
  AC match {
    case Some(ac) => ac
    case None => Normal(V)
  }
```

**Figure 13.** The `while` statement in the IR interpreter

The implementation of the interpreter is still in progress. Currently, it does not support the `eval` code, and it partially supports the `with` statement, the strict mode, and the standard built-in objects. We are actively developing the missing pieces, and we are validating the interpreter implementation using regression testing.

### 4.4 Experiences with the Quirky Semantics

Because the ECMAScript specification does not provide a high-level overview of the quirky semantics of JavaScript in one place, we have encountered various inconsistencies and under-specified behaviors sporadically. We share some of such experiences here.

Unlike conventional programming languages where the value of a non-empty sequence of statements is the value of the last statement of the sequence regardless of the value, in JavaScript, it is the value of the last statement that produces a non-`empty` value, if any; otherwise, the value of the sequence is `empty`. For example, in the following code, an excerpt from the ECMAScript specification [9]:

```
eval("1;;;;;")
eval("1;{}")
eval("1;var a;")
```

the calls to the `eval` function all return the value `1`. If we do not consider this semantics and translate an AST to an IR as we do for the conventional languages, the translation will not preserve the semantics. At the same time, if we apply such a semantics to get the value of a sequence of IR statements, it will not preserve the semantics either because a single AST statement often gets translated into a sequence of IR statements; a single AST statement of the value `empty` can be translated into a sequence of IR statements whose values are not all `empty`. For example, while the following JavaScript code:

```
if (false) { 42; }
```

produces the value empty, the translated IR from the code:

```
<>_temp_1 = false;
if (<>_temp_1) { <>_temp_2 = 42; };
```

consists of IR statements whose values are false and empty, in order. Thus, we should treat a sequence of IR statements translated from a single AST statement differently from a sequence of IR statements translated from a sequence of AST statements. For this reason, we introduce the IRStmtUnit node in addition to the IRSeq node: we wrap a sequence of IR statements translated from a single AST statement with IRStmtUnit, and we use IRSeq to wrap a sequence of IR statements translated form a sequence of AST statements. Thus, the value of a sequence of IR statements wrapped by IRStmtUnit is the value of the last statement in the sequence, while the value of a sequence of IR statements wrapped by IRSeq is the last non-empty value of the sequence.

The flexibility of JavaScript allows the number of arguments to a function call to be different from the number of the function's parameters. When the number of the arguments is smaller than the number of the parameters, the parameters whose corresponding arguments do not exist are mapped to undefined. More specifically, evaluation of the arguments at a function call creates an Arguments object, which contains properties of the names from "0" to the maximum number of the number of the arguments, say $n_a$, and the number of the parameters, say $n_p$. Each of the properties maps to the corresponding values of the arguments, except that when $n_a$ is smaller than $n_p$ the properties of the names from $n_a$ to $n_p - 1$ map to undefined.

In JavaScript, many standard built-in objects are functions in the sense that they can be invoked with arguments, and at the same time, some of them are constructors in the sense that they can be used with the new operator. The issue here is that the semantics of some built-in objects are very different depending on whether they are used as functions or as constructors. For example, when a Date object is used as a function[9, Section 15.9.2], it ignores its arguments and behaves as it is called as (new Date()).toString() returning a string value, while it creates a Date object when it is used as a constructor[9, Section 15.9.3]. To handle this semantics correctly, our AST-to-IR translation creates separate IR nodes for function calls and object constructions even though their semantics are very similar.

Because JavaScript provides prototype-based inheritance, every object has an internal property called [[Prototype]]. Also, an object may have a prototype of name "prototype"; in particular, every function automatically has a prototype property so that it can be used as a constructor [9, Section 8.6.2]. A somewhat confusing part is that while [[Prototype]] and prototype are quite different, they are used in an intermingled way. Because [[Prototype]] is an internal property, users can not access it explicitly and the value of the property does not change once it is set when the object is constructed. However, users can explicitly access and change the value of prototype of an object freely. In spite of such differences, they are used in an intermixed way which may lead to some confusion. For instance, a instanceof b denotes whether b has a as its instance, and the ECMAScript specification defines its semantics as the result of calling the [[HasInstance]] internal method of b with the argument of a, which checks whether the value of the "prototype" property of b exists in the [[Prototype]] chain from the [[Prototype]] internal property of a. Because users can change the value of the "prototype" property of b any time, the result of the instanceof operation can be surprising as we have shown in Figure 1 in Section 1.

Another issue is *implementation-dependent* and *implementation-defined* features in the ECMAScript specification. For example, the description of Array.prototype.sort [9, Section 15.4.4.11] includes several implementation-dependent semantics. When an Array object includes some missing elements, it is said to be *sparse*, and the semantics of Array.prototype.sort is full of implementation-dependent behavior for sparse arrays. While the evaluation of the Array.prototype.sort method involves evaluating the [[Get]], [[Put]], and [[Delete]] internal methods, it is often impossible to evaluate them for sparse arrays. If a sparse array has the [[Extensible]] internal property of false, it is impossible to put a non-existing array index property, and if a sparse array has an index property with the [[Configurable]] internal property of false, it is impossible to delete the property. Such implementation-dependent and implementation-defined semantics make it difficult to analyze JavaScript programs.

## 5. Related Work

The first thorough approach to formally specify a significant (if not all) portion of the JavaScript programming language is an operational semantics [23] that directly translates the ECMAScript specification. It intends to faithfully capture the various aspects of JavaScript and translate the informal ECMAScript specification into a formal semantics. Therefore, it is well suited for analyzing existing JavaScript programs and applying the analysis results back to the JavaScript programs, though reasoning about the programs and proving their properties using the semantics are not conventional and complicated [11].

Another approach for a formal specification of JavaScript is to devise a core calculus, $\lambda_{JS}$ [17], which has a very different semantics from the original JavaScript semantics but more conventional. By providing a set of desugaring (or rather, compilation) rules from JavaScript to $\lambda_{JS}$, it desugars away the quirky features of JavaScript and translates them to a standard core calculus with a conventional semantics. Thus, it suits well for extending and reasoning about the JavaScript language thanks to its conventional calculus, though extra work is necessary to scale the analysis results back to the full JavaScript language.

Jensen *et al.* [20] present a static analysis framework for JavaScript using type analysis for the full JavaScript language. Their framework is originally based on the T. J. Watson Libraries for Analysis (WALA) [34], which is a collection of Java libraries to support static analyses of Java bytecode and JavaScript. Because WALA originally supported only Java programs, its support for JavaScript is yet premature. It parses JavaScript programs using the Rhino parser, and it translates the parsed JavaScript AST in Rhino into the WALA common AST. Thus, it has the same problem as Rhino regarding supporting any research that modifies the JavaScript syntax. Instead, Jensen *et al.* developed their own framework to support the peculiar features of JavaScript more directly.

Recent research on JavaScript covers various aspects of the JavaScript language such as its formal specifications [11, 17, 23], type systems [3, 18, 33], static analyses [16, 20], and combinations of static and dynamic analyses [5, 24] for JavaScript. Implementation of such research results often requires extensive modifications of an existing framework. Implementing a proposed JavaScript module system [21] requires modifications in the parser and the AST structure, for example. Based on our experience, the handwritten Rhino parser is too complicated to be applicable, and the big semantic gap between JavaScript and $\lambda_{JS}$ and poor performance of the $\lambda_{JS}$ interpreter are obstacles to be usable. We believe that our framework would be useful for easily implementing future research on JavaScript.

## 6. Conclusion

Motivated by our own struggles with the existing approaches, we present SAFE, a scalable analysis framework for ECMAScript, the very first attempt to support both formal specification and implementation of JavaScript, developed open for the research community. To make the framework more usable to other researchers, SAFE provides not only the formal specification of its various components but also the implementation of them. Designing and developing the formal specification and implementation of the JavaScript language simultaneously has been a worthwhile approach; the operational semantics rules in our formal specification and the pattern matching mechanism in our implementation checked by the Scala compiler have been mutually beneficial to catch missing cases in each other. Also, thorough investigation of the ECMAScript specification revealed several issues and error-prone semantics of JavaScript, which opens up possibilities for future research in detecting erroneous programs. In order for the framework to be as flexible, scalable, and pluggable as possible, SAFE supports three levels of intermediate representations: AST, the closest to the JavaScript source, IR for evaluation, and CFG for flow-based analyses. Last but not least, we actively use third-party open-source tools to automatically generate parsers and intermediate representations, which enhances the productivity of aggressive research and development on JavaScript.

Currently, we have been finalizing the implementation of the interpreter with extensive testing, and we have been collaborating with researchers in academia and industry. As we have shown in Figure 2 in Section 2.2, such collaborative research includes `CloneDetector`, `Coverage`, and `Analyzer`. Each research topic performs on different intermediate representations: `CloneDetector` detects possible code clones among multiple JavaScript programs via AST comparisons, `Coverage` calculates code coverage of JavaScript programs via IR evaluation, and `Analyzer` estimates type information via CFG traversal. We plan to improve our current analyzer and add more analyses to the framework, and we more than welcome other researchers using our framework for their own research. As a theoretical foundation of our JavaScript semantics specification, we're working on mechanizing the formal semantics in Coq to guarantee some properties. While we're supporting most of the built-in objects both in the interpreter implementation and an accompanying analysis implementation, we do not intend to formally describe built-in objects and functions exhaustively.

## Acknowledgments

The authors sincerely thank the S-Core team for the fruitful discussions and their tremendous contributions to the framework. This work is supported in part by the Mid-career Researcher Program of Korea Ministry of Education, Science and Technology(MEST) / National Research Foundation of Korea(NRF) (Grant 2012-0005256), the Engineering Research Center of Excellence Program of MEST / NRF (Grant 2012-0000469), Microsoft Research Asia, Samsung Electronics, and S-Core.

## References

[1] HTML5: A vocabulary and associated APIs for HTML and XHTML. `http://www.w3.org/TR/html5/`.

[2] Alexa the Web Information Company. `http://www.alexa.com/`, 2011.

[3] Christopher Anderson, Paola Giannini, and Sophia Drossopoulou. Towards type inference for JavaScript. In *Proceedings of the 19th European Conference on Object-Oriented Programming*, 2005.

[4] Gilad Bracha, Guy Steele, Bill Joy, and James Gosling. *Java$^{TM}$ Language Specification, The 3rd Edition (Java Series)*. Addison-Wesley Professional, 2005.

[5] Ravi Chugh, Jeffrey A. Meister, Ranjit Jhala, and Sorin Lerner. Staged information flow for JavaScript. In *Proceedings of the 2009 ACM SIGPLAN Conference on Programming Language Design and Implementation*, 2009.

[6] Douglas Crockford. ADsafe. `http://www.adsafe.org/`.

[7] Brendan Eich. Static analysis FTW. `http://brendaneich.com/2010/08/static-analysis-ftw/`, 2010.

[8] European Association for Standardizing Information and Communication Systems (ECMA). ECMA-262: ECMAScript Language Specification. Third edition., 1999.

[9] European Association for Standardizing Information and Communication Systems (ECMA). ECMA-262: ECMAScript Language Specification. Fifth edition., 2009.

[10] Facebook. FBJS. `http://developers.facebook.com/docs/fbjs/`.

[11] Philippa Anne Gardner, Sergio Maffeis, and Gareth David Smith. Towards a program logic for javascript. In *Proceedings of the 39th Annual ACM SIGPLAN-SIGACT Symposium on Principles of Programming Languages*, 2012.

[12] Danny Goodman. *Dynamic HTML: The Definitive Reference*. O'Reilly Media, 1998.

[13] Google. Caja. `http://code.google.com/p/google-caja/`.

[14] Google. V8 javascript engine. `http://code.google.com/p/v8/`.

[15] Robert Grimm. Rats! – An Easily Extensible Parser Generator. `http://cs.nyu.edu/~rgrimm/xtc/rats.html`.

[16] Arjun Guha, Shriram Krishnamurthi, and Trevor Jim. Using static analysis for Ajax intrusion detection. In *Proceedings of the 18th International Conference on World Wide Web*, 2009.

[17] Arjun Guha, Claudiu Saftoiu, and Shriram Krishnamurthi. The essence of JavaScript. In *Proceedings of the 24th European Conference on Object-Oriented Programming*, 2010.

[18] Phillip Heidegger and Peter Thiemann. Recency types for analyzing scripting languages. In *Proceedings of the 24th European Conference on Object-Oriented Programming*, 2010.

[19] Dongseok Jang, Ranjit Jhala, Sorin Lerner, and Hovav Shacham. An empirical study of privacy-violating information flows in JavaScript web applications. In *Proceedings of the 17th ACM Conference on Computer and Communications Security*, 2010.

[20] Simon Holm Jensen, Anders Møller, and Peter Thiemann. Type analysis for JavaScript. In *International Static Analysis Symposium*, 2009.

[21] Seonghoon Kang and Sukyoung Ryu. Formal specification of a JavaScript module system. In *Proceedings of the 2012 ACM International Conference on Object Oriented Programming, Systems, Languages, and Applications (OOPSLA '12)*, 2012.

[22] Brian W. Kernighan and Dennis M. Ritchie. *The C Programming Language*. Prentice Hall, 1988.

[23] Sergio Maffeis, John C. Mitchell, and Ankur Taly. An operational semantics for JavaScript. In *Proceedings of the 6th Asian Symposium on Programming Languages and Systems*, 2008.

[24] Sergio Maffeis, John C. Mitchell, and Ankur Taly. Isolating JavaScript with filters, rewriting, and wrappers. In *14th European Symposium on Research in Computer Security*, 2009.

[25] Sergio Maffeis, John C. Mitchell, and Ankur Taly. Object capabilities and isolation of untrusted web applications. In *IEEE Symposium on Security and Privacy*, 2010.

[26] Mozilla. Rhino: Javascript for java. `https://developer.mozilla.org/en-US/docs/Rhino`.

[27] Mozilla. Spidermonkey. `https://developer.mozilla.org/en-US/docs/SpiderMonkey`.

[28] Tim O'Reilly. What Is Web 2.0. `http://oreilly.com/web2/archive/what-is-web-20.html`.

[29] Changhee Park, Hongki Lee, and Sukyoung Ryu. An empirical study on the rewritability of the with statement in JavaScript. In *2011 International Workshop on Foundations of Object-Oriented Languages (FOOL'11)*, 2011.

[30] KAIST PLRG. JSAF: JavaScript Analysis Framework. `http://plrg.kaist.ac.kr/research/safe`.

[31] Joe Gibbs Politz, Spiridon Aristides Eliopoulos, Arjun Guha, and Shriram Krishnamurthi. Type-based verification of JavaScript sandboxing. In *Proceedings of the 20th Conference on USENIX Security Symposium*, 2011.

[32] Brian R. Stoler, Eric Allen, and Dan Smith. ASTGen. `http://sourceforge.net/projects/astgen`.

[33] Peter Thiemann. Towards a type system for analyzing JavaScript programs. In *Proceedings of the 14th European Symposium on Programming*, 2005.

[34] T.J. Watson Libraries for Analysis (WALA). `http://wala.sf.net`.